\documentclass[11pt]{article}
\usepackage[utf8]{inputenc}
\usepackage{geometry}
\usepackage{graphicx}
\usepackage{algorithm}
\usepackage{algpseudocode} 
\usepackage{float}
\usepackage{authblk}
\usepackage{titlesec}
\usepackage{hyperref}
\usepackage{caption}
\usepackage{amsmath}
\usepackage{booktabs}
\usepackage{appendix} 
\usepackage{url} 
\titleformat{\section}{\large\bfseries}{\thesection.}{1em}{}
\titleformat{\subsection}{\normalsize\bfseries}{\thesubsection.}{1em}{}

\title{\textbf{Performance Evaluation of Efficient Hybrid Compression Methods for Devanagari-Encoded Hindi Text Using Lossless Algorithms}}
\author[1]{Mukesh Sahu}
\author[2]{Jeebananda Panda}
\affil[1]{Delhi Technological University, Delhi 110042, India}
\affil[2]{Delhi Technological University, Delhi 110042, India}

\date{}  

\begin{document}

\maketitle

\begin{abstract}
This research paper provides a comprehensive performance analysis of five standard lossless compression algorithms—LZMA, Zstd, Brotli, Bzip2, and LZ4HC—alongside sixty hybrid combinations on UTF-8 encoded Hindi text datasets, written in the Devanagari script, across varying sizes—small, medium, and large. The evaluation covers compression ratio, compression speed, decompression speed, and a derived weighted normalized efficiency score. The hybrid combination Zstd + LZ4HC achieves the highest efficiency, with scores of 0.6764 for small files and 0.8597 for large files. The analysis underscores the significance of combining fast and high-ratio compression techniques, particularly for language-specific data.
\end{abstract}


\section{Introduction}
Data Compression is an important issue for digital communication system. Data as it is essential for minimizing storage requirements and enhancing transmission efficiency, particularly for languages like Hindi. Hindi text, encoded using the Devanagari script and represented in Unicode, typically results in larger file sizes due to the presence of multi-byte characters. These characteristics lead to irregular character frequency patterns, making conventional compression algorithms less effective.\\
Given Hindi’s widespread usage across government, education, and media sectors, developing effective compression methods is crucial—especially in regions with bandwidth or storage constraints. Better compression directly improves data accessibility, lowers infrastructure costs, and supports the broader dissemination of Hindi content in digital form.\\
This study focuses on hybrid compression techniques that apply two lossless algorithms sequentially. Specifically, we examine the following five algorithms: LZMA (Lempel-Ziv-Markov chain algorithm), Zstd (Zstandard), Brotli, Bzip2, and LZ4HC (High Compression variant of LZ4). These algorithms were selected due to their proven performance in text compression and their ability to balance high compression ratios with fast processing speeds.\\
•	Zstd is recognized for its fast compression and decompression speeds, making it suitable for real-time applications.[1]\\
•	Brotli, developed by Google, has demonstrated strong performance in web compression, providing a balance between speed and compression ratio.[2]\\
•	Bzip2, with its Burrows-Wheeler transform (BWT), is known for its excellent compression ratio but at the cost of slower compression speeds, making it ideal for cases where compression efficiency is prioritized over speed. [3,4]\\
•	. LZMA, known for its high compression ratio, has been widely used in applications where space-saving is critical, particularly for archiving and large dataset [5]\\
•	LZ4HC is a high-compression variant of LZ4, providing fast decompression speeds while achieving a higher compression ratio, suitable for scenarios where quick access to compressed data is required.[6]\\
By combining these algorithms in hybrid configurations, we aim to leverage the strengths of each, addressing the unique challenges posed by the Devanagari script. This approach not only seeks to achieve efficient compression but also ensures fast processing times, making it feasible for real-world applications where both storage and speed are of paramount importance.

\section{Literature Review}
The literature reveals extensive research in the field of lossless text compression, with focus areas ranging from standard algorithm evaluations to language-specific and hybrid approaches.

\subsection{Standard Compression Algorithms}

Salomon and Motta [7] provide a foundational overview of data compression algorithms, highlighting trade-offs in compression ratio, speed, and complexity—establishing theoretical grounding for hybrid methods. Jayasankar et al. [8] conducted a survey on data compression (DC) techniques, categorizing them based on data quality, coding schemes, data types, and applications. They provided a comparative analysis and highlighted open research challenges for future development.Fauzan et al. [9] evaluate Huffman and LZW algorithms for lossless compression, finding LZW more efficient for .txt and .csv files with space savings of 63.85 percent and 77.56 percent. Stecuła et al. [10] compared the performance of standard compression algorithms (zlib, lzma, bz2, zl4) across languages such as Esperanto, Polish, and English. Their study demonstrated that Esperanto provided better compression results than both Polish and English, highlighting the influence of language characteristics on compression efficiency. Ignatoski et al. [11] compared LZW and arithmetic coding across eight languages, finding that language structure—especially alphabet size and text type—significantly affects compression efficiency, with LZW generally outperforming. The study underscores the need for language-aware compression strategies Indurani et al. [12] reviewed standard compression methods like GZ, BZIP2, SMAZ, and SHOCO for short text in big data, emphasizing their role in efficient storage and transmission. Nasif et al. [13] review classical algorithms like Huffman and LZ77, including their hybrid form Deflate, underlining their relevance in cloud computing. However, these studies do not explore the compression behavior of Indic scripts or address the need for language-specific adaptations. Gupta et al. [14] compare Deflate, Bzip2, LZMA, PPMd, and PPMonstr using the Silesia corpus, providing valuable benchmarking across general-purpose datasets. Yet , the evaluation remains limited to English and lacks insights into language-based optimization for scripts like Devanagari.
\subsubsection{Hybrid Compression Techniques} 
Jrai et al. [15] introduced a multi-level preprocessing with adaptive LZW for Arabic Unicode text, achieving a 71 percent compression rate and outperforming standard methods like LZW, DEFLATE, and Gzip. The study underscores the effectiveness of language-specific preprocessing. Baidoo [16] compares Huffman, Arithmetic, RLE, and LZW on text files of varying complexity. The study finds that compression efficiency varies by algorithm and file type, highlighting the importance of selecting algorithms based on compression goals. Tariq et al. [17] propose serial combinations of algorithms to improve compression outcomes through sequential hybridization. Abdulmonim and Muhamad [18] proposed a hybrid compression model combining Huffman, RLE, and LZW, achieving a 2.42 average compression ratio and improved file size metrics compared to individual techniques. This supports the effectiveness of sequential hybrid methods in enhancing lossless text compression. Silué et al. [19] compare various lossless compression techniques (Huffman, LZW, RLE, BWT) on a custom text dataset. The results show that LZW achieves the best compression ratio, while hybrid methods like BWT+RLE also perform well, highlighting the importance of selecting algorithms based on data characteristics. A limitation of these works is the absence of testing on Unicode-based or non-English text corpora, which restricts its applicability to multilingual contexts
\subsection{Indic Language Compression} 
Gupta et al. [20] address Devanagari composite character compression using a hybrid of Huffman and run-length coding, achieving up to 67 percent compression savings. However, their method is primarily applied to font-level pattern compression and is not generalized to full-text Hindi datasets encoded in Unicode. Maniya et al. [21] explore a dictionary-based compression technique tailored for Gujarati text. It compares the performance of Huffman coding and other compression methods on UTF-8 encoded Gujarati text, highlighting the effectiveness of dictionary-based approaches. Vijayalakshmi and Sasirekha [22] developed a Tamil text compression approach outperforming traditional schemes like Huffman, ZIP, and LZW. Yet, the methodology is tailored specifically to Tamil and does not explore extensibility to other Indic languages. Addepalli and Lakshmi [23] applied LZW-based compression within a Hadoop framework to handle large-scale medical data.
While, their focus was on infrastructure rather than linguistic characteristics or language-aware optimization.
Amaeer et al. [24] introduced a Unicode-reduction scheme using 8-bit transformation for compressing Bangla natural text. Although effective for Bangla, the approach has not been adapted for other Indic languages or evaluated within hybrid configurations. Aslanyürek and Mesut [25] proposed WSDC, a static dictionary-based method enhanced by k-means clustering for short text compression. However, their framework does not include Devanagari script and primarily focuses on multilingual but non-specific corpora.

While the above mentioned studies provide valuable insights into hybrid compression techniques, there is a noticeable gap in research specifically targeting Hindi text compression. The unique characteristics of the Devanagari script, including its complex ligatures and diacritics, pose challenges that are not directly addressed by existing hybrid methods. This gap highlights the need for developing adaptive, customizable hybrid compression frameworks tailored to the linguistic and structural nuances of Hindi text.hybrid compression for Hindi remains an underexplored area.

\section{Methodology}

A total of 75 algorithm–dataset combinations were evaluated in this study, encompassing five standard compression algorithms—LZMA, Zstd, Brotli, Bzip2, and LZ4HC—and their various hybrid configurations. Each algorithm or hybrid was applied to three UTF-8 encoded Hindi text datasets of different sizes (small, medium, and large) to facilitate a comprehensive comparative analysis.
\subsection{Experimental Setup}
The experiments were conducted on a system equipped with an AMD Ryzen 9 5900HX processor clocked at 3.30 GHz, with 16 GB DDR4 RAM, running Windows 11 Home (64-bit). All scripts and evaluations were performed using Python 3.11 (64-bit) within the IDLE environment.
All source code, datasets, and replication instructions used in this study are publicly available on GitHub [30].
\subsection{Compression Datasets} 
Three UTF-8 encoded Hindi text datasets of varying sizes were used for evaluation. The small dataset is approximately 145 KB, the medium dataset around 1,600 KB, and the large dataset roughly 13,000 KB. These represent short-term, medium-term, and long-term text data scenarios, respectively.
\subsection{Implementation Details} 
The compression framework employs a sequential hybrid approach, where two compression algorithms are applied in succession. Standard configurations were used for all algorithms: LZMA was set to level 6 with no dictionary; Zstd also used level 6 with no dictionary; Brotli was configured with quality level 6 and a window size of 22; Bzip2 was executed with its default settings and a block size of 900 KB; and LZ4HC was applied at level 6 without a dictionary.
\subsection{Hybrid Compression-Decompression Workflow} 
To support the hybrid framework, each algorithm combination follows a two-step process during compression and decompression. The pseudocode below illustrates the simplified workflow applied uniformly across all algorithm–dataset combinations in this study

\begin{figure}[htbp]
    \centering
    \includegraphics[width=0.5\textwidth]{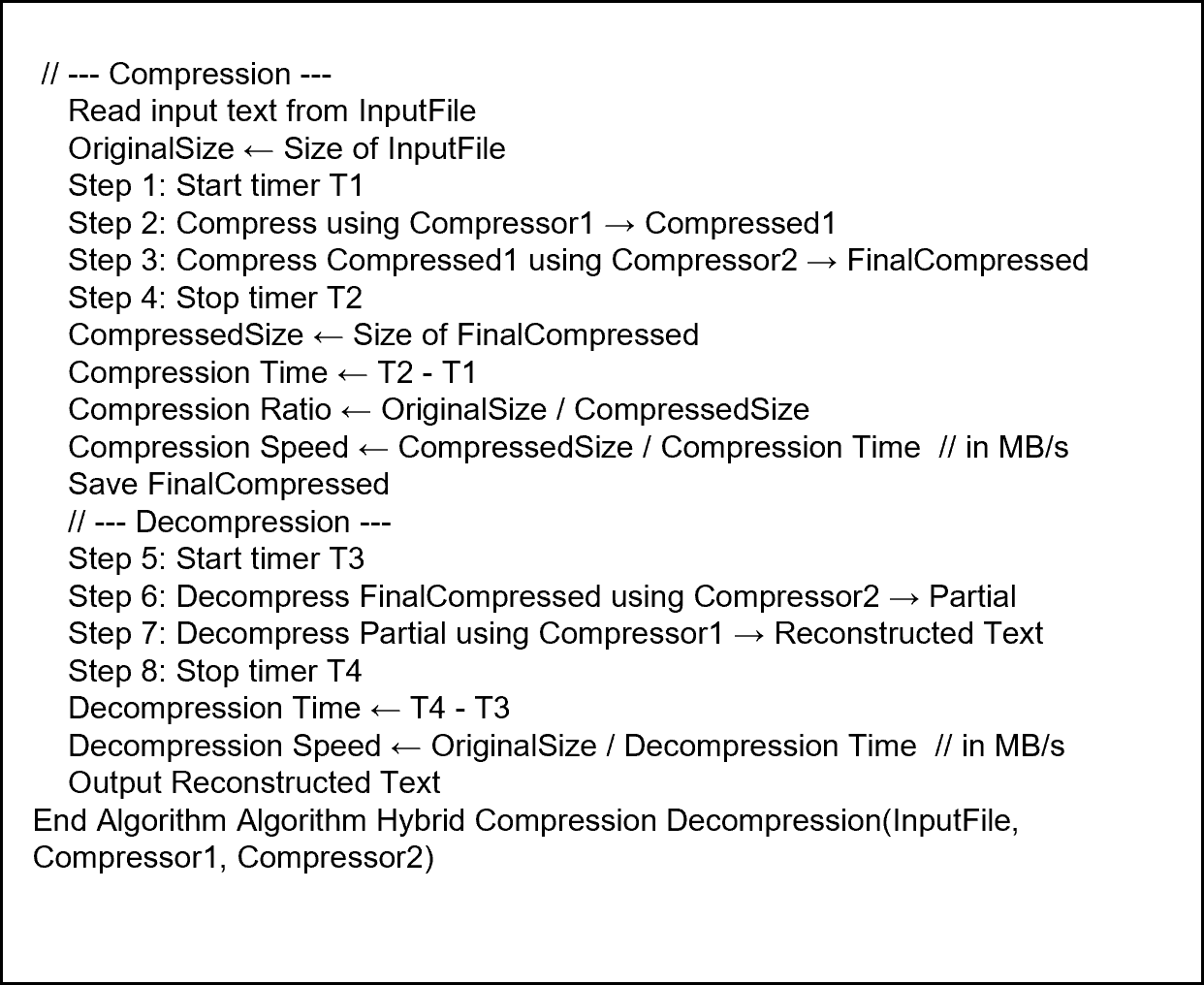} 
    \caption{ Hybrid Compression and Decompression with Metric Calculation
} 
    \label{Algorithm 1.} 
\end{figure}

\subsection{Performance Metrics} 
\subsubsection{Compression Ratio (CR)} 
The compression ratio is calculated by dividing the original size of the file by its compressed size:
\begin{equation}
\mathrm{CR} = \frac{\text{Original File Size (in kilobytes)}}{\text{Compressed File Size (in kilobytes)}}
\label{Equation 1:compression_ratio}
\end{equation}

This metric indicates how effectively an algorithm reduces the file size. A higher value signifies better compression efficiency.
\subsubsection{Compression Speed (CS)} 
The compression speed is determined by dividing the compressed size by the time taken for compression		
\begin{equation}
\mathrm{CS} = \frac{\text{Original File Size (in MB)}}{\text{Time Taken to Compress (in seconds)}}
\label{Equation 2:compression_speed}
\end{equation}

Measured in megabytes per second (MB/s), this reflects how fast the algorithm compresses the data. Higher values denote faster performance.
\subsubsection{ Decompression Speed (DS)}
The decompression speed is calculated by dividing the original size by the decompression time
\begin{equation}
\mathrm{DS} = \frac{\text{Compressed File Size (in MB)}}{\text{Time Taken to Decompress (in seconds)}}
\label{Equation 3 :decompression_speed}
\end{equation}

Also in MB/s, this indicates the speed at which compressed data is restored. Higher values are desirable.
\subsubsection{Formula and Efficiency Calculation}
Compression efficiency (E) is evaluated using min-max normalized metrics:
\begin{equation}
E = (0.40 \times \text{CR}_{\text{norm}}) + (0.30 \times \text{CS}_{\text{norm}}) + (0.30 \times \text{DS}_{\text{norm}})
\label{Equation 4:efficiency}
\end{equation}

where compression ratio (CR), compression speed (CS), and decompression speed (DS) are normalized as:
\begin{equation}
X_{\text{norm}} = \frac{X - X_{\text{min}}}{X_{\text{max}} - X_{\text{min}}}
\label{Equation 5:normalization}
\end{equation}

Normalization ensures fair comparison. The 40\% + 30\% + 30\% 
 weighting prioritizes compression ratio (40 percent) for better storage efficiency, while compression and decompression speeds (30 percent each) ensure computational feasibility. This balanced evaluation strategy aligns with previous comparative studies on lossless data compression algorithms that emphasize compression ratio, speed, decompression speed and efficiency as key performance indicators [26–29].

\section{Results and Discussion}
This section presents a detailed evaluation of the compression efficiency of six standard independent algorithms and their hybrid combinations (75 algorithms-dataset combinations) when applied to Hindi text files of varying sizes. The analysis is conducted on three datasets: small, medium, and large files. The results are benchmarked using four parameters: compression ratio (Equation 1), compression speed ( equation 2), decompression speed (Equation 3) and a derived efficiency metric. Efficiency was computed using a composite score: 40 percent compression ratio, 30 percent compression speed, and 30 percent decompression speed (Equation 4 and 5).

\subsection{ Compression Ratio}
\begin{table}[H]
\centering
\caption{Comparison of Independent and Hybrid Algorithms in Terms of Compression Ratio}
\begin{tabular}{@{}p{3cm}p{6cm}p{6cm}@{}}
\toprule
\textbf{Dataset} & \textbf{Independent Algorithm} & \textbf{Best Hybrid} \\ \midrule
Small File       & Bzip2 (6.57)                   & Bzip2 + Zstd (6.57)  \\
Medium File      & Bzip2 (7.56)                   & Bzip2 + Brotli (7.6) \\
Large File       & Brotli (117.11)                & LZMA + Brotli (141.59) \\
\bottomrule
\end{tabular}
\end{table}

According to table 1 for small files, both the independent algorithm Bzip2 (6.57) and the hybrids Bzip2 + Brotli / Zstd (6.57) offered the highest compression ratio, showing that some hybrids can match the best-performing standalone compressors .For medium files, the hybrid Bzip2 + Brotli (7.6) marginally outperformed independent Bzip2 (7.56).For large files, the hybrid LZMA + Brotli (141.59) significantly exceeded Brotli(117.11) by offering 21 percent improvement in compression ratios, the best independent algorithm—demonstrating the effectiveness of deep-layered compression for larger datasets
\subsection{Compression Speed}
\begin{table}[H]
\centering
\caption{Comparison of Independent and Hybrid Algorithms in Terms of Compression Speed}
\begin{tabular}{@{}p{3cm}p{5cm}p{5cm}@{}}
\toprule
\textbf{Dataset} & \textbf{Independent Algorithm}\\ {(Speed MB/s)} & \textbf{Hybrid Algorithm }\\{(Speed MB/s)} \\
\midrule
Small File       & Zstd (45.99)          & Zstd + LZ4HC (46.31) \\
Medium File      & Zstd (76.61)          & Zstd + Brotli (70.34) \\
Large File       & Zstd (546.22)         & Zstd + Brotli (1071.57) \\
\bottomrule
\end{tabular}
\end{table}

Table 2 shows that Zstd led among standalone algorithms for all file sizes.While, the hybrid Zstd + LZ4HC (46.31 MB/s) slightly exceeded Zstd (45.99 MB/s) for small files, while Zstd + Brotli (1071.57 MB/s) more than doubled Zstd's speed for large files, making it ideal for high-throughput compression applications
\subsection{Decompression Speed} 

\begin{table}[H]
\centering
\caption{Comparison of Independent and Hybrid Algorithms in Terms of Decompression Speed}
\begin{tabular}{@{}p{3cm}p{6cm}p{6cm}@{}}
\toprule
\textbf{Dataset} & \textbf{Independent Algorithm (Speed MB/s)} & \textbf{Hybrid Algorithm (Speed MB/s)} \\ \midrule
Small File       & LZMA (281.19)         & LZ4HC + Zstd (323.83) \\
Medium File      & LZMA (456.98)         & LZ4HC + Zstd (485.04) \\
Large File       & Zstd (593.79)          & Zstd + LZ4HC (663.78) \\
\bottomrule
\end{tabular}
\end{table}

Table 3 illustrates that for small and medium files, the hybrid \textbf{LZ4HC + Zstd} outpaced 
    
\subsection{ Compression Efficiency}
Efficiency was computed using a  normalized composite score: 40 percent compression ratio, 30 percent compression speed, and 30 percent decompression speed (Equation 4 and 5).

\begin{table}[H]
\centering
\caption{Comparison of Independent and Hybrid Algorithms in Terms of Compression Efficiency}
\begin{tabular}{@{}p{3cm}p{6cm}p{6cm}@{}}
\toprule
\textbf{Dataset} & \textbf{Independent Algorithm (Efficiency Score)} & \textbf{Hybrid Algorithm (Efficiency Score)} \\ \midrule
Small File       & Zstd (0.5406)         & Zstd + LZ4HC (0.6764) \\
Medium File      & Zstd (0.6293)         & Zstd + Brotli (0.582) \\
Large File       & Zstd (0.6836)         & Zstd + LZ4HC (0.8567) \\
\bottomrule
\end{tabular}
\end{table}

 From table 4 it is evident that results clearly highlight the dominance of hybrid algorithms, particularly combination like Zstd + LZ4HC in large files which outperform standard Zstd by nearly 25.7 percent and provides optimal compression without compromising speed

\section{Statistical and Comparative Analysis}
\subsection{Trends in Compression Efficiency Across Dataset Sizes} 

\begin{table}[H]
\centering
\caption{Top 10 Compression Algorithms Based on Weighted and Normalized Efficiency Scores for Small, Medium, and Large Hindi Text Datasets}
\normalsize 
\resizebox{\textwidth}{!}{  
\begin{tabular}{@{}c|l|c|l|c|l|c@{}}
\toprule
\textbf{Rank} & \textbf{Algorithm/Hybrid
(Small Files)} & \textbf{Efficiency (Small files)} & \textbf{Algorithm/Hybrid (Medium Files)} & \textbf{Efficiency (Medium files)} & \textbf{Algorithm/Hybrid (Large Files)} & \textbf{Efficiency(Large files)} \\ \midrule
1 & Zstd + LZ4HC & 0.6764 & Zstd (Independent) & 0.6293 & Zstd + LZ4HC & 0.8597 \\
2 & LZ4HC + Zstd & 0.5708 & Zstd + Brotli & 0.582 & Zstd + Brotli & 0.8057 \\
3 & Bzip2 + LZ4HC & 0.5591 & Zstd + LZ4HC & 0.5678 & Zstd (Independent) & 0.6836 \\
4 & Bzip2 + Zstd & 0.5556 & Bzip2 + Zstd & 0.4908 & Zstd + Bzip2 & 0.6779 \\
5 & Bzip2 + Brotli & 0.5476 & Bzip2 + Brotli & 0.4907 & Brotli + Zstd & 0.6683 \\
6 & Brotli + Zstd & 0.5430 & Bzip2 + LZ4HC & 0.4899 & Zstd + LZMA & 0.6539 \\
7 & Zstd (Independent) & 0.5406 & LZ4HC + Zstd & 0.482 & Brotli + LZ4HC & 0.6116 \\
8 & Zstd + Brotli & 0.4951 & Bzip2 (Independent) & 0.4717 & Brotli (Independent) & 0.6073 \\
9 & LZMA (Independent) & 0.4839 & Bzip2 + LZMA & 0.4584 & Brotli + LZMA & 0.5817 \\
10 & Bzip2 + LZMA & 0.4788 & Brotli + Zstd & 0.462 & Brotli + Bzip2 & 0.573 \\
\bottomrule
\end{tabular}
}  
\end{table}

\begin{figure}[htbp]
    \centering
    \includegraphics[width=0.8\textwidth]{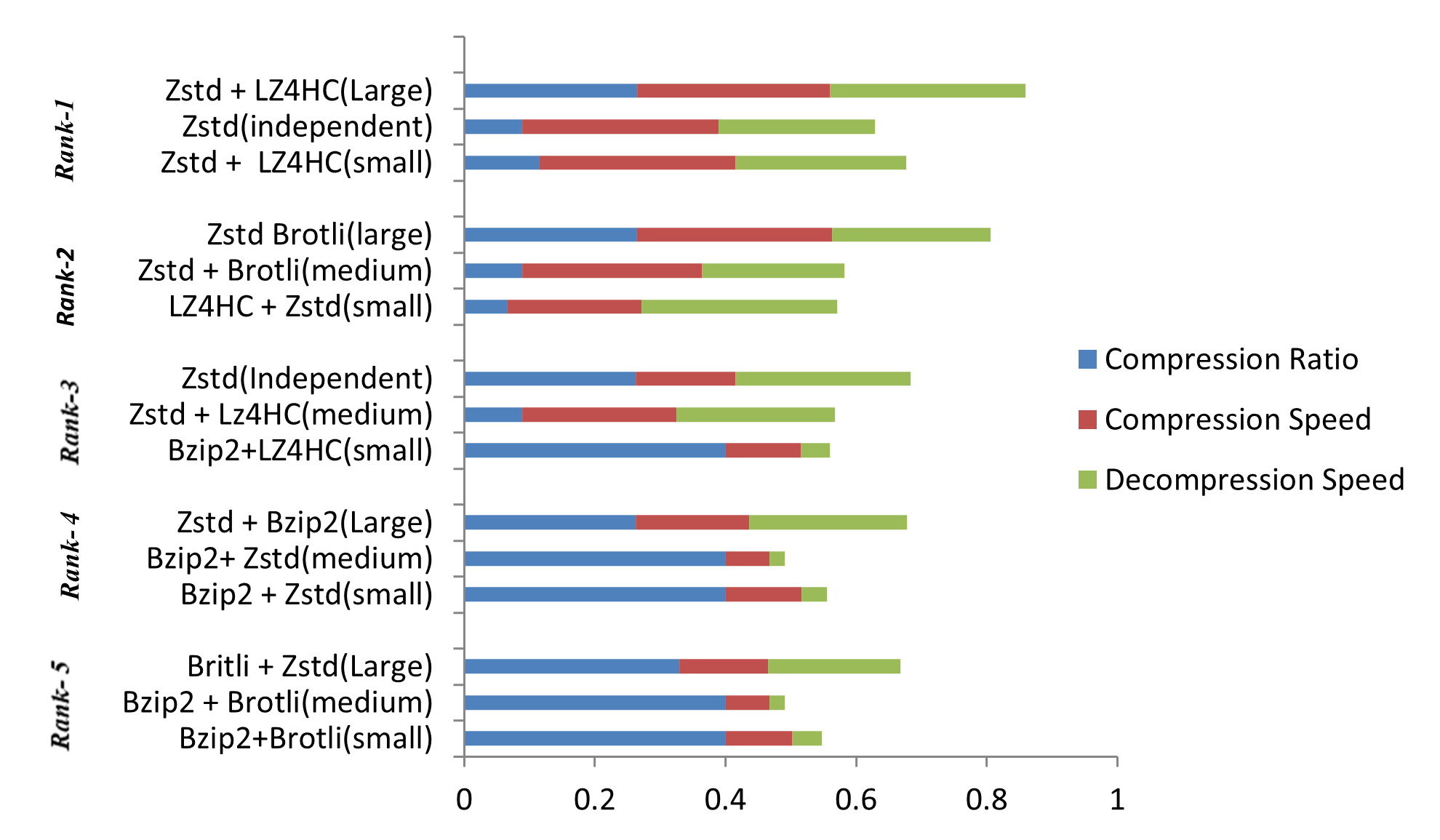} 
    \caption{Normalized and Weighted Efficiency Comparison of Compression Techniques for Hindi Text} 
    \label{ Normalized and Weighted Efficiency Comparison of Compression Techniques for Hindi Text} 
\end{figure}

Table 5 and Figure 2 jointly illustrate the comparative efficiency of the top ten compression algorithms and hybrid methods across three dataset sizes—small, medium, and large. A consistent trend emerges wherein hybrid combinations significantly outperform most standalone algorithms, particularly for larger datasets. The hybrid method Zstd + LZ4HC secures the highest efficiency scores for both small (0.6764) and large (0.8597) files, while the standalone Zstd achieves the top rank for medium-sized files (0.6293). The bar chart in Figure 1 visually reinforces this pattern, highlighting the dominance of Zstd-centric hybrid configurations such as LZ4HC + Zstd, Bzip2 + Zstd, and Zstd + Brotli across various ranks and file sizes. Notably, the visual layout also emphasizes the increasing contribution of decompression speed (green bars) in the overall efficiency of top-ranking hybrids, particularly for larger datasets. This integrated analysis suggests that hybrid models leveraging Zstd as a core component consistently yield superior performance due to their ability to balance high compression ratio, fast compression speed, and efficient decompression. Consequently, Zstd + LZ4HC stands out as the most balanced and adaptable compression technique across diverse Hindi text data volumes.
To further validate the efficiency of the proposed hybrid algorithm, a direct performance comparison between Zstd + LZ4HC and the independent standard compression algorithms was conducted on the large file dataset. This comparison highlights the hybrid’s ability to maintain a high compression ratio while significantly improving speed, resulting in superior overall efficiency.
\subsection{Performance Comparison of the Proposed Hybrid Algorithm}
\begin{table}[H]
\centering
\caption{Performance Comparison of the Proposed Hybrid Algorithm (Zstd + LZ4HC) with Standard Compression Algorithms on Large Hindi Text Files}
\resizebox{\textwidth}{!}{%
\begin{tabular}{@{}lcccc@{}}
\toprule
\textbf{Algorithm} & \textbf{Compression Ratio} & \textbf{Compression Speed (MB/s)} & \textbf{Decompression Speed (MB/s)} & \textbf{Efficiency Score} \\ \midrule
Zstd + LZ4HC  & 94.82  & 1055.69 & 663.78 & 0.8597 \\
Zstd (Independent) & 94.49 & 546.22 & 593.79 & 0.6836 \\
Brotli (Independent) & 117.11 & 312.66 & 426.47 & 0.6073 \\
Bzip2 (Independent) & 9.77 & 18.75 & 4.88 & 0.021 \\
LZ4HC (Independent) & 3.79 & 32.51 & 554.34 & 0.258 \\
LZMA (Independent) & 141.91 & 5.96 & 16.56 & 0.0573 \\
\bottomrule
\end{tabular}
}
\end{table}

Table 6 presents this comparative analysis, showcasing the balanced advantage that Zstd + LZ4HC offers over all standard methods.. The hybrid achieves a strong efficiency score of 0.8597, combining a high compression ratio (94.82) with exceptional speed (1055.69 MB/s compression, 663.78 MB/s decompression). Among standard algorithms, Zstd performs best independently (efficiency: 0.6836) but remains outpaced by the hybrid. Others like LZMA and Bzip2, though offering high compression ratios, suffer from slow speeds, resulting in very low efficiency (0.0573 and 0.021, respectively). This concise comparison underscores the effectiveness of hybridization in balancing speed and compression quality for large-scale Hindi text compression.

\subsection{Performance Balance Analysis (Compression Ratio vs. Compression Speed)} 

 We have analyze compression ratio vs. compression speed for large files, since that's where trade-offs are most pnounced
\begin{table}[ht]

\centering
\caption{Performance Balance Analysis}
\begin{tabular}{|l|c|c|}
\hline
\textbf{Algorithm} & \textbf{Compression Ratio} & \textbf{Compression Speed (MB/s)} \\
\hline
LZMA + Brotli & 141.59 & 6.09 \\
LZMA + LZ4HC & 141.10 & 6.01 \\
Zstd + Brotli & 94.49 & 1071.57 \\
Zstd + LZ4HC & 94.82 & 1055.69 \\
Brotli + Zstd & 117.10 & 491.50 \\
Brotli + LZ4HC & 117.08 & 426.05 \\
Zstd (indep.) & 94.59 & 546.22 \\
Brotli + LZMA & 117.5 & 208.44 \\
\hline
\end{tabular}
\end{table}

\begin{figure}[htbp]
    \centering
    \includegraphics[width=0.8\textwidth]{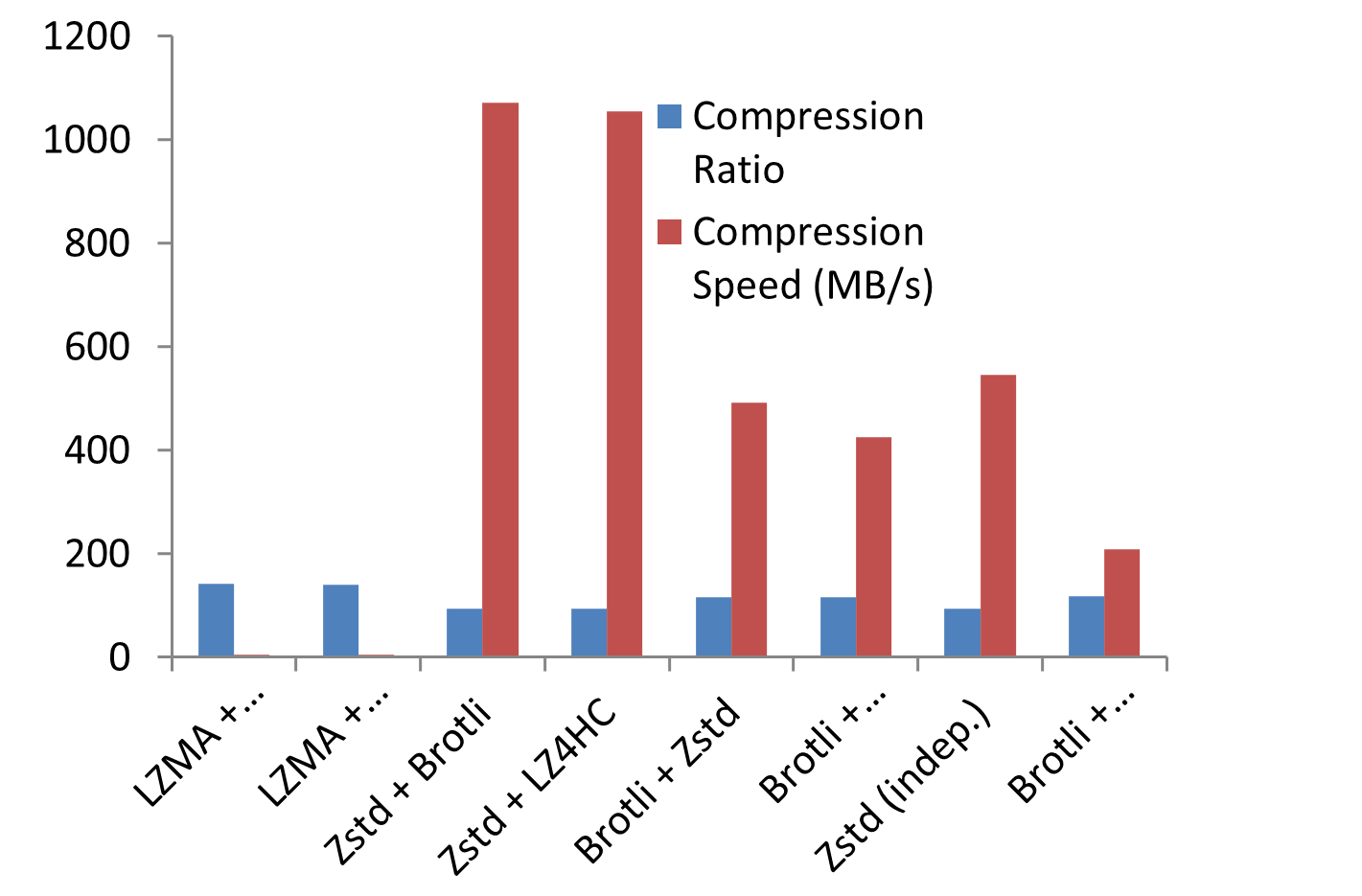} 
    \caption{Compression Ratio vs. Compression Speed Analysis} 
    \label{fig:graph2} 
\end{figure}

Table 7 and figure 3. analysis reveals that Zstd-based hybrids consistently achieve the best performance balance for large files. In particular, Zstd + Brotli and Zstd + LZ4HC stand out by offering exceptionally high compression speeds (over 1000 MB/s) while maintaining competitive compression ratios (~94.5). These are well-suited for real-time or high-throughput scenarios.
On the other hand, combinations like LZMA + Brotli and LZMA + LZ4HC deliver the highest compression ratios (above 141) but at significantly reduced speeds, making them ideal for storage-focused tasks where time is not critical.
Hybrid algorithms involving Brotli (e.g., Brotli + Zstd, Brotli + LZ4HC) emerge as strong all-rounders, balancing both compression ratio and speed effectively.
Overall, the results affirm that hybrid approaches can be tailored to specific needs, outperforming many standalone algorithms by strategically combining the strengths of two components.

\subsection{Contribution Frequency of Core Algorithms in Hybrid Compression Techniques
}  
   \begin{figure}[htbp]
    \centering
    \includegraphics[width=0.8\textwidth]{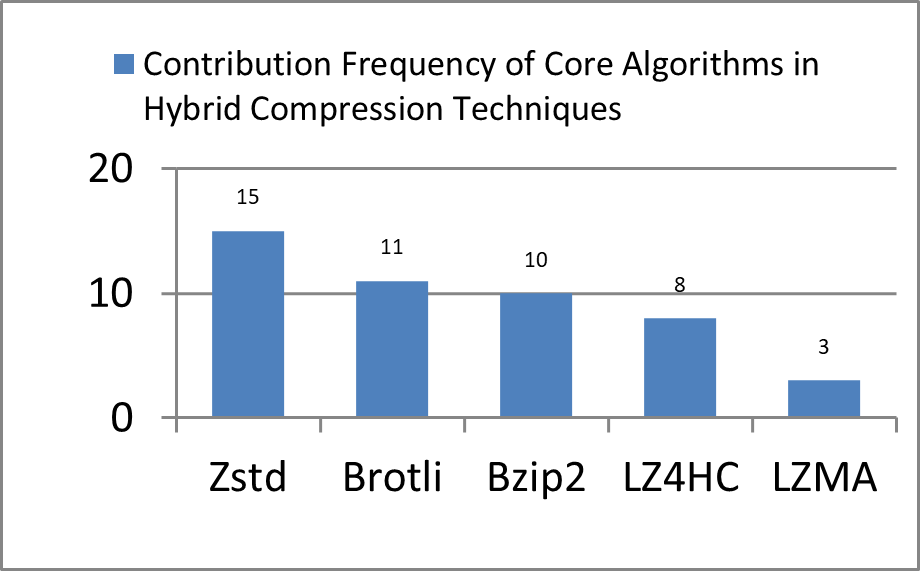}
    \caption{ Frequency of Top Components in Top-Performing Hybrids. Total combinations analyzed: 30 (10 per size category)}
    \label{FIGURE 4. }
\end{figure}
              
Contribution Frequency of Core Algorithms in Hybrid Compression Techniques
 Figure 4 shows the component impact analysis of the frequency  in the top-performing hybrid compression algorithms reveals distinct trends in their relative contribution to overall efficiency. Among the 30 combinations evaluated across small, medium, and large Hindi text files, Zstd emerged as the most frequently occurring component, appearing in 15 of the top hybrids—accounting for 50 percent of all top-performing combinations. This was followed by Brotli, which appeared 11 times (36.7percent), and Bzip2, included in 10 instances (33.3percent). LZ4HC showed a solid presence with 8 appearances (26.7percent), while LZMA appeared less frequently, in only 3 of the top hybrids (10 percent). These trends suggest that Zstd, Brotli, and Bzip2 are more likely to contribute to balanced and high-efficiency compression outcomes, making them strong candidates for constructing optimized hybrid algorithms. The dominance of Zstd in particular underscores its adaptability and performance when paired with other compressors.

\section{Conclusion}
This research highlights the remarkable potential of hybrid compression techniques for efficiently compressing Devanagari-encoded Hindi text, a script known for its complexity and richness. By evaluating a range of lossless algorithms, both independent and hybrid, across diverse datasets, this study uncovers key insights into how these methods perform under various compression and decompression conditions.
The results underscore a clear trend: hybrid algorithms consistently outperform their standard counterparts in terms of compression ratio, speed,decompression speed and overall performance. Among the various combinations tested, the hybrid pairing of Zstd + LZ4HC(Among top 3 across all the data sets) emerged as the standout performer, offering a 93.27 percent improvement in compression speed, 11.8 percent better decompression speed, and ~25 percent higher weighted and normalized efficiency compared to standard algorithms like Zstd in large dataset files. striking the optimal balance between compression ratio and processing time. This demonstrates that leveraging the strengths of multiple algorithms offers a clear advantage, particularly for large-scale text datasets.
These findings have profound implications for the digital preservation and transmission of Hindi text, as well as for developing applications in areas like machine learning, natural language processing, and web services. The results demonstrate that by optimizing compression for Devanagari-encoded Hindi, we can significantly improve the efficiency of handling large volumes of Hindi data in digital environments. Future research could focus on refining these hybrid models to further enhance performance, ensuring that such methods continue to evolve alongside the growing need for efficient multilingual data compression.
\subsection{Real-World Applications } 
The hybrid compression techniques presented in this study offer significant economic and societal benefits. By reducing the file sizes of Hindi text, these methods can lower storage costs and decrease the bandwidth needed for data transmission. This is especially valuable for digital libraries, archival systems, and online content delivery platforms, which often operate under budget and connectivity constraints. , the improved compression efficiency enhances accessibility to regional language content. Faster data transfer and reduced storage requirements can make digital educational resources, government documents, and cultural content more widely available, thereby supporting digital inclusion and literacy across diverse populations
\subsection{Implications and Future Work}
Although the current study demonstrates that hybrid algorithms—particularly the combination of Zstd + LZ4HC—consistently outperform standard methods in terms of compression ratio, speed, and overall efficiency, our findings also suggest that there is room for further optimization. In particular, the experimental results indicate that the performance of hybrid compression techniques varies with dataset size and text characteristics.
Based on these observations, we propose that future research should explore the development of an adaptive, customizable compression framework. Such a system would dynamically select or configure compression algorithms based on the specific properties of the Hindi text and the performance requirements of the application. By tailoring the compression strategy—whether to prioritize storage efficiency, processing speed, or a balanced trade-off—this adaptive framework could significantly enhance the practical application of compression techniques for Devanagari-encoded Hindi text. We believe that further investigation in this area could lead to a dedicated research study, which would build upon the findings of the current work.

\appendix  

\section{Appendix A: Data and Code Availability}
The source code, Hindi text datasets, and performance evaluation results used in this study
are available at the public GitHub repository: \texttt{Sahusir. Hybrid Compression Algorithms for
Devanagari Text. GitHub. Available at: \url{https://github.com/Sahusir/Hybrid-Compression-Algorithms-for-Devanagari-Text}}.

The repository is structured as follows:

\begin{itemize}
    \item \textbf{code/}: Python scripts implementing hybrid compression pipelines.
    \item \textbf{data/}: UTF-8 encoded Hindi text datasets (small, medium, large).
    \item \textbf{result/}: Output files and compressed data for each experiment.
    \item \textbf{paper/}: Supplementary files including appendix and metrics.
\end{itemize}

A comprehensive \texttt{README.md} file provides replication steps, environment setup instructions, and details on input/output formats to facilitate reproducibility.

\end{document}